\documentstyle[12pt]{article}
\normalbaselines
\begin{document}

\baselineskip=24pt

\bibliographystyle{unsrt}
\vbox{\vspace{6mm}}

\begin{center}
{\Large Interference effects in f-deformed 
fields}
\end{center}

\bigskip
\bigskip
\bigskip

\begin{center}
{\it Stefano Mancini $^{\dag}$, Vladimir I. Man'ko $^{\ddag}$
and Paolo Tombesi $^{\dag}$}
\end{center}

\bigskip

\begin{center}
$^{\dag}$ Dipartimento di Matematica e Fisica, 
Universit\`a di Camerino,
I-62032 Camerino, Italy\\
Istituto Nazionale di Fisica Nucleare, 
Sezione di Perugia, Italy\\
and Istituto Nazionale di Fisica della Materia, 
Unit\`a di Camerino, Italy
\end{center}

\begin{center}
$^{\ddag}$ Lebedev Physical Institute, Leninsky Pr. 53,
117924 Moscow, Russia
\end{center}

\bigskip
\bigskip
\bigskip

\begin{abstract}
We show how the introduction of an algeabric field 
deformation affects 
the interference phenomena. We also give a physical 
interpretation of the 
developed theory.
\end{abstract}


\bigskip
\bigskip
\bigskip

Nonlinear systems have always attracted a lot of attention both in 
classical and in quantum fields. On the other hand, after their 
introduction 
\cite{bied}, also the quantum q-oscillators were interpreted 
\cite{Napoli} as a 
nonlinear oscillators with a very specific type of nonlinearity, 
in which 
the frequency of vibration depends on the energy of these 
vibrations, 
through an hyperbolic cosine function containing the parameter of 
nonlinearity. But there might exsist other types of nonlinearity 
for which 
the frequency of oscillations varies  with the amplitude  by means 
of a 
generic function, say $f$. For this reason, recentely, it has been 
introduced the concept of f-deformed oscillators \cite{scripta}.
The particular case of f-coherent states, called also nonlinear 
coherent 
states for the function $f$, expressed in terms of Laguerre 
polynomials 
has been shown reachable in trapped ions \cite{vogel}.

The specific and important property of any linear process is 
the existence
of the superposition principle, due to which two solutions
of the linear equation may superpose and give rise to another 
solution of the
linear equation. Physically it means the possibility of the 
interference 
phenomenon when two different solutions, with appropriate phases 
both in time and in space domains, produce a stable pattern 
corresponding
to increasing and cancelling amplitudes of both solutions in 
concrete points of space (time).
If the equation has some nonlinearity there is no 
superposition principle anymore, but if the nonlinearity is 
small it is 
clearly 
intuitive that the interference pattern, characteristics for 
purely
linear vibrations, will be only slightly changed (deformed) 
according to
the influence of the nonlinearity.
The nonlinearity may produce generation of other harmonics, 
which may 
each other interact, 
implying the beating phenomenon. 
In th time domain it could mean a collapse and revival of the
interference pattern, as well as in the space domain it could mean 
the existence of
a spoty structure, with sharp enough pattern picture in one 
spot and with
a different pattern in another
spot. 
The influence of the nonlinearity into interference patterns 
may be
traced out by using the influence of the nonlinearity onto 
the visibility, 
which is a characteristic of coherence properties of
the waves under study.

Here we would discuss the coherence properties of f-deformed fields 
showing how their deformation could affect the visibility of the 
interference pattern. 
We also provide a physical realization of the
developed theory within the context of the Bose-Einstein  
condensate.

Let us start by considering two fields described by the 
annihiliation (creation) 
operators $\hat a$ (${\hat a}^{\dag}$) and $\hat b$ 
(${\hat b}^{\dag}$).
The free fields Hamiltonian, in the case of unit frequencies 
(setting $\hbar=c=1$),
will be
\begin{equation}\label{H}
{\hat H}=\frac{1}{2}\left({\hat a}{\hat a}^{\dag}
+{\hat a}^{\dag}{\hat a}\right)
+\frac{1}{2}\left({\hat b}{\hat b}^{\dag}
+{\hat b}^{\dag}{\hat b}\right)\,,
\end{equation}
which yields an operators evolution in the Heisemberg 
picture of the form
\begin{equation}\label{aboft}
{\hat a}(t)={\hat a}e^{-it}\,;\quad
{\hat b}(t)={\hat b}e^{-it}\,.
\end{equation}
The interference effects arise as a consequence of fields 
superposition, 
which leads to a total field of the form
\begin{equation}\label{Psi}
{\hat\Psi}(t)=\left[{\hat a}(t)e^{-ik_1x}
+{\hat b}(t)e^{-ik_2x}\right]+h.c.\,,
\end{equation}
where $k_1$ ($k_2$) is the momentum characterizing the 
field $a$ ($b$).
Then, the total field intensity will be
\begin{equation}\label{I}
I(x,t)=\langle{\hat\Psi}^{(-)}(t){\hat\Psi}^{(+)}(t)\rangle
=\left\langle\left[{\hat a}^{\dag}(t)e^{ik_1x}
+{\hat b}^{\dag}(t)e^{ik_2x}\right]
\left[{\hat a}(t)e^{-ik_1x}+{\hat b}(t)e^{-ik_2x}\right]
\right\rangle\,.
\end{equation}
Due to the fact that the two initial fields are indipendent,
the interference fringes appear only if we take the expectation 
value over 
particular states of the two modes, like the coherent states
\begin{equation}\label{abstates}
|\alpha\rangle_a\otimes|\beta\rangle_b\,,
\end{equation}
obtaining
\begin{equation}\label{Inondef}
I(x,t)=2|\alpha|^2\{1+\cos[\phi(x)-\phi]\}\,,
\end{equation}
where for semplicity we have choosen $\beta=\alpha e^{-i\phi}$ 
and we 
have set $\phi(x)=(k_1-k_2)x$. 
The visibility of the interference fringes is given by 
\cite{qnoise}
\begin{equation}\label{Vdef}
{\cal V}=\frac{I(x,t)_{\rm max}-I(x,t)_{\rm min}}
{I(x,t)_{\rm max}+I(x,t)_{\rm min}}\,,
\end{equation}
hence, in Eq. (\ref{Inondef}) it results as the factor 
multiplying
the cosine term inside the brackets, i.e.
${\cal V}=1$.

Let us now introduce  a deformation of the fields by means of the 
following Hamiltonian \cite{scripta}
\begin{equation}\label{Hcal}
{\hat{\cal H}}=\frac{1}{2}\left({\hat A}{\hat A}^{\dag}
+{\hat A}^{\dag}{\hat A}\right)
+\frac{1}{2}\left({\hat B}{\hat B}^{\dag}
+{\hat B}^{\dag}{\hat B}\right)\,,
\end{equation}
where 
\begin{equation}\label{AB}
{\hat A}={\hat a}f({\hat n}_a,{\hat n}_b)\,;\quad
{\hat B}={\hat b}f({\hat n}_a,{\hat n}_b)\,,
\end{equation}
with $f$ a generic function of the operators 
${\hat n}_a={\hat a}^{\dag}{\hat a}$ and
${\hat n}_b={\hat b}^{\dag}{\hat b}$. 
It should be remarked that in this form one introduces 
a coupling between 
the two modes $a$ and $b$, other than a self interaction 
of the fields.

By using the commutation properties of the fields operators, the 
Hamiltonian (\ref{Hcal}) can be rewritten as 
\begin{equation}\label{Hcalnew}
{\cal H}({\hat n}_a,{\hat n}_b)=
\frac{1}{2}\left[({\hat n}_a+{\hat n}_b)f^2({\hat n}_a,{\hat n}_b)
+({\hat n}_a+1)f^2({\hat n}_a+1,{\hat n}_b)
+({\hat n}_b+1)f^2({\hat n}_a,{\hat n}_b+1)\right]\,,
\end{equation}
which determines the following time evolution of the operators
\begin{eqnarray}
{\hat a}(t)&=&{\hat a}\exp\left\{
-i\left[{\cal H}({\hat n}_a,{\hat n}_b)
-{\cal H}({\hat n}_a-1,{\hat n}_b)\right]t\right\}\,;
\label{aoftdef}\\
{\hat b}(t)&=&{\hat b}\exp\left\{
-i\left[{\cal H}({\hat n}_a,{\hat n}_b)
-{\cal H}({\hat n}_a,{\hat n}_b-1)\right]t\right\}\,.
\label{boftdef}
\end{eqnarray}
Thus, inserting Eqs. (\ref{aoftdef}) and (\ref{boftdef}) 
into Eq. (\ref{I}) and using 
the states of Eq. (\ref{abstates}), we get
\begin{equation}\label{Idef}
I(x,t)=2|\alpha|^2\left[1+
\Re\left\{e^{-2|\alpha|^2}
\sum_{n_a,n_b=0}^{\infty}\frac{|\alpha|^{2(n_a+n_b)}}{n_a!n_b!}
e^{i[{\cal H}(n_a,n_b+1)-{\cal H}(n_a+1,n_b)]t}
e^{i[\phi-\phi(x)]}\right\}\right]\,.
\end{equation}
If we further suppose to have the function $f$ symmetric 
under the exchange
${\hat n}_a\leftrightarrow{\hat n}_b$, then the visibility becomes
\begin{equation}\label{Vdef}
{\cal V}=e^{-2|\alpha|^2}
\sum_{n_a,n_b=0}^{\infty}\frac{|\alpha|^{2(n_a+n_b)}}{n_a!n_b!}
e^{i[{\cal H}(n_a,n_b+1)-{\cal H}(n_a+1,n_b)]t}\,,
\end{equation}
which clearly shows the time dependence through the specific 
function $f$.

Alternatively, one can consider the two mode deformed fields 
not entangled, 
i.e.
\begin{equation}\label{ABnotent}
{\hat A}={\hat a}f_a({\hat n}_a)\,;\quad
{\hat B}={\hat b}f_b({\hat n}_b)\,,
\end{equation}
in this case the modified Hamiltonian is given by
\begin{equation}\label{Hnotent}
{\cal H}({\hat n}_a,{\hat n}_b)=
\frac{1}{2}\left[{\hat n}_af_a^2({\hat n}_a)
+{\hat n}_bf_b^2({\hat n}_b)
+({\hat n}_a+1)f_a^2({\hat n}_a+1)
+({\hat n}_b+1)f_b^2({\hat n}_b+1)\right]\,,
\end{equation}
and the visibility takes the same form of Eq. (\ref{Vdef})
provided to have $f_a=f_b$.

We now apply the above arguments to atom optics, where 
an increasing interest has been devoted to the Bose-Einstein 
condensates 
after their observation \cite{and}.
First, we take
\begin{equation}\label{f1cond}
f^2_a({\hat n}_a)=\kappa{\hat n}_a+(1-\kappa)\,;\quad
f^2_b({\hat n}_b)=\kappa{\hat n}_b+(1-\kappa)\,,
\end{equation}
which gives an Hamiltonian of the type
\begin{equation}\label{H1cond}
{\cal H}({\hat n}_a,{\hat n}_b)={\hat n}_a+{\hat n}_b
+\kappa\left({\hat n}_a^2+{\hat n}_b^2\right)\,.
\end{equation}
It could describe two trapped condensates each with self 
collisional 
effects, with $\kappa$ representing the collisional rate 
between the atoms 
within each condensate \cite{Walls}. 
In this case the argument of the exponential in Eq. (\ref{Vdef}) 
takes the form $2i\kappa(n_b-n_a)t$ leading to collapses and 
revivals of the 
visibility \cite{Walls}.

Second, we take
\begin{equation}\label{f2cond}
f^2({\hat n}_a,{\hat n}_b)=\kappa\left({\hat n}_a
+{\hat n}_b\right)
+(1-\kappa)\,,
\end{equation}
which gives an Hamiltonian of the type
\begin{equation}\label{H2cond}
{\cal H}({\hat n}_a,{\hat n}_b)
={\hat n}_a+{\hat n}_b
+\kappa\left({\hat n}_a+{\hat n}_b\right)^2\,,
\end{equation}
which could describe two trapped condensates, including cross 
collisional 
effects as well \cite{tony}.
Of course one could consider a cross collisional rate 
different from 
$\kappa$, but it will result a function $f$ no longer 
symmetric in the 
exchange ${\hat n}_a\leftrightarrow{\hat n}_b$, and for
semplicity we 
will not take into account this eventuality here.
 It is easy to see that in the particular case of 
 Eq. (\ref{H2cond}), 
 the visibility 
does not show time dependence. 
It means that the correlation between the two modes, 
induced by the Hamiltonian (\ref{H2cond}), provides 
to maintain the initial 
visibility.

In conclusion, we have investigated the superposition 
mechanism of 
f-deformed fields showing that the formalism of deformed 
oscillators 
could result a powerfull tool in atom optics.

\end{document}